\documentclass{article}
\usepackage[utf8]{inputenc}
\usepackage{times}  % DO NOT CHANGE THIS
\usepackage{helvet} % DO NOT CHANGE THIS
\usepackage{courier}  % DO NOT CHANGE THIS
\usepackage[hyphens]{url}  % DO NOT CHANGE THIS
\usepackage{graphicx} % DO NOT CHANGE THIS
\usepackage{graphicx}  % DO NOT CHANGE THIS
\usepackage{cleveref}
\newcommand{\myf}{\fontfamily{lmtt}\selectfont}
\newcommand{\blue}{\color{blue}}
\newcommand{\red}{\color{red}}
\usepackage{graphicx}
\usepackage{colortbl}
\usepackage{adjustbox}
\usepackage{subfigure}
\title{The Effects of Gender Signals and Performance in Online Product Reviews}
\author{Sandipan Sikdar\textsuperscript{\rm 1}\footnote{sandipan.sikdar@cssh.rwth-aachen.de}, Rachneet Singh Sachdeva\textsuperscript{\rm 1}, Johannes Wachs\textsuperscript{\rm 1},\\ Florian Lemmerich\textsuperscript{\rm 1} and Markus Strohmaier\textsuperscript{\rm 1}\textsuperscript{\rm 2}\\
\textsuperscript{1}RWTH Aachen University, Germany\\
\textsuperscript{2}GESIS, Cologne, Germany}
\begin{document}

\maketitle
\begin{abstract}
This work quantifies the effects of signaling and performing gender on the success of reviews written on the popular amazon.com shopping platform. Highly rated reviews play an important role in e-commerce since they are prominently displayed below products. Differences in how gender-signaling and gender-performing review authors are received can lead to important biases in what content and perspectives are represented among top reviews. 
To investigate this, we extract signals of author gender from user names, distinguishing reviews where the author's likely gender can be inferred. Using reviews authored by these gender-signaling authors, we train a deep-learning classifier to quantify the gendered writing style or gendered performance of reviews written by authors who do not send clear gender signals via their user name. We contrast the effects of gender signaling and performance on review success using matching experiments. While we find no general trend that gendered signals or performances influence overall review success, we find strong context-specific effects. For example, reviews in product categories such as  {\myf Electronics} or {\myf Computers} are perceived as less helpful when authors signal that they are likely woman, but are received as more helpful in categories such as {\myf Beauty} or {\myf Clothing}. In addition to these interesting findings, our work provides a general chain of tools for studying gender-specific effects across various social media platforms. 
\end{abstract}

\section{Introduction}

Differences in social outcomes between genders have been studied in many diverse online contexts including Wikipedia~\cite{wagner2015s}, social media~\cite{nilizadeh2016twitter}, online newspapers~\cite{jia2016women}, and online freelance marketplaces~\cite{hannak2017bias}. Despite this previous work, it is still unclear how differences in outcomes relate to responses to signals of gender identity or to the performance of gender in content. In this work, we analyze how users of the popular online shopping platform Amazon.com rate the helpfulness of online product reviews depending on gendered signals and performance of their authors. Given that these ratings impact the visibility of reviews and hence their influence on product success, gender bias may shape the market in unexpected ways~\cite{boyd2012critical}.

\noindent\textbf{Research goals}: We first aim to understand the effects of explicit signals of a reviewer's gender on the reception of their product review. For that purpose, we compare content by users signaling a likely gender in their user name to the case when no reliable inference about an author's gender can be made. To bridge the two we infer the gendered performance of all reviewers, including those who do not signal their likely gender via their user names, using data extracted from the text of the reviews. 

In particular, our analysis focuses on two specific research questions. First, is there a difference in the success of product reviews, measured by ratings of ``helpfulness'' made by other users, depending on the likely gender of the author? Second, is there an effect on the appreciation of reviews if the gender of an author is signaled or performed?

\begin{figure*}
\centering
\includegraphics[scale=0.37]{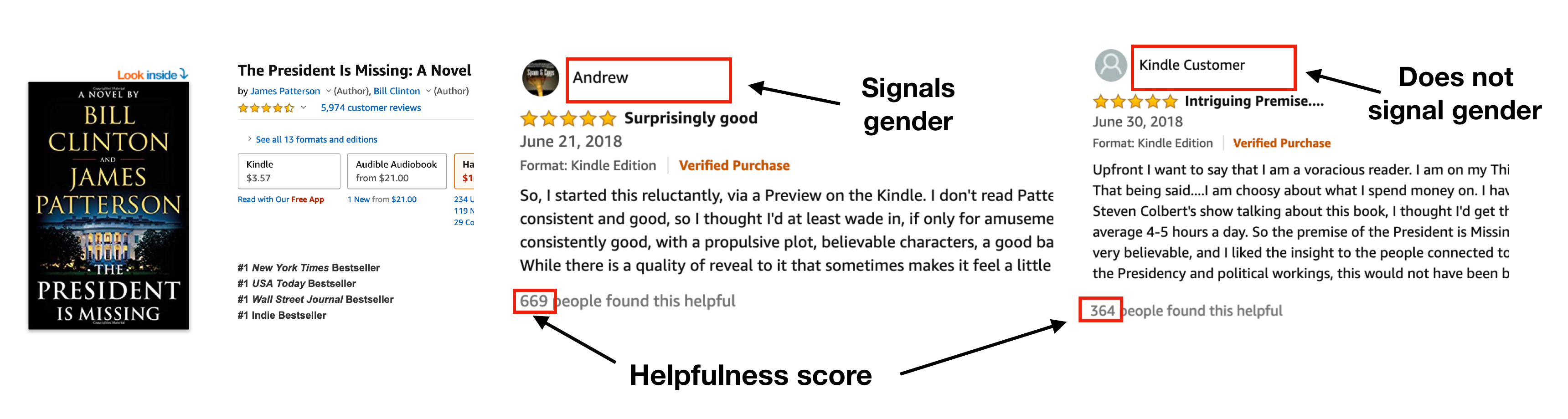}
\caption{\label{fig_1} \emph{Illustration:} Consider the reviews written for the product `The President is Missing: A Novel'. The author of the review on the left signals a likely gender by the user name 'Andrew'. On the right the author's user name `Kindle Customer' does not signal a likely gender. Further, the perceived helpfulness of a review can be quantified by the helpfulness score of the review. We explore whether signals of the gender of the authors influences the perceived helpfulness of a review.} 
\end{figure*}

To illustrate our setting, we show two example reviews in Figure~\ref{fig_1}. In the review on the left we observe a signal of the author's likely gender via his user name (``Andrew''). In contrast, we cannot reasonably infer the likely gender of the author of the review on the right.
Other users on Amazon can express appreciation for a review by marking it as helpful. The number of such appraisals of a review, called its ``helpfulness score,'' is displayed below each review. We investigate if there is a relationship between user gender signals and the helpfulness score the review receives. If there is a relationship between gender disclosure and feedback, how does it vary across the kinds of products reviewed on Amazon?

\noindent\textbf{Approach and methods}: We study a dataset with more than 80 million reviews written by around $21$ million users about ca. 9 million products on `Amazon.com'\footnote{https://www.amazon.com}, Amazon's North American domain. Using the name-to-gender prediction tool \emph{Gender-Guesser}\footnote{https://pypi.org/project/gender-guesser/}, we are able to label the gender of reviewers for 42\% of the total reviews in the dataset that are in line with human judgment in $98\%$ of cases.

With these signaled gender labels, we employ character-level Convolutional Neural Networks (CNN) as state-of-the-art text classification methods to measure the gendered behavior or performance of reviews that do not signal gender. We can infer the signaled gender of a user in a held-out test set with an overall accuracy 82\% using the text features. Using this measure of performed gender, we categorize user reviews into four groups: (i) signaling men, (ii) signaling women, (iii) performing men, (iv) performing women. We then perform a set of matching experiments to compare the helpfulness scores of reviews in the different groups in otherwise (e.g., with respect to publication time, review length, sentiment, product rating, etc.) similar reviews and study differences in social feedback these users receive across product categories.

\noindent\textbf{Results}: 
We find that on average across all the categories, gender signaling does not have an effect on the perceived helpfulness of product reviews. However, we observe substantial context-specific effects. For example, reviews authored by signaling women, are perceived as more helpful in products related to categories like {\myf Movies} or {\myf Beauty}. By contrast, we notice increased helpfulness scores for signaling men in categories such as {\myf Electronics} or {\myf Kindle}. Comparing signaling women with performing women, we find that signaling gender hurts in categories such as {\myf Electronics}, {\myf Games} or {\myf Computers}. Similar negative effects are observed for signaling vs. performing men in product categories including {\myf Clothing}, {\myf Beauty} or {\myf Toys}. We consider the presence of such context-sensitive effects as a primary finding of this paper that significantly extends previous findings in the field.
Our results promote increased awareness of gender-specific effects in the perception of online reviews and suggest implications for online platforms (e.g., regarding the ranking of reviews or display of user names if they potentially imply gender) and their users (e.g., awareness of such effects when choosing a user name).

\section{Dataset Preparation}
In this section, we describe the data we use in this study and introduce the methods we employ, to approach our research questions. In particular, we describe how gender of reviewers can be inferred from user names and review content. 

\subsection{Dataset}
We leverage a publicly available dataset of Amazon product reviews\footnote{\url{http://jmcauley.ucsd.edu/data/amazon/}, "aggressively deduplicated data" version} consisting of reviews written between May 1996 and July 2014.\cite{he2016ups,mcauley2015image}
Each review contains information on the author's user name, the product rating (between $1$ and $5$ stars), a helpfulness score (i.e., the number of users who marked the review as helpful), the reviewed product, the date of the review, and its text. Each product is linked to meta data including a description of the product, category information (each product can belong to multiple categories), price, brand, and image features. 

The dataset contains about $80$ million reviews (excluding around $2$ million reviews with missing attributes) of $20.9$ million unique reviewers about $9.01$ million products assigned to $18.1$ thousand categories.  On average, each review contains $84.8$ words and rates the product with $4.16$ stars. Regarding helpfulness, reviews receive on average $2.07$ upvotes and $0.71$ downvotes. Information on the gender is not directly available from this dataset, and we will discuss the methods for inferring gender signals in the following section.

\subsection{Gender signals though user names}
\label{disclose_gender}
We now describe how we infer the perceived, signaled gender of reviewers. In general inferring the gender of online users is known to be a challenging task~\cite{karimi2016inferring,lin2016recognizing}. To alleviate this problem, we first make the simplifying assumption of binary gender~\ref{sec:limitations}.
We therefore identify users as likely (signaling) men or likely (signaling) women by applying the name-based gender prediction tool \emph{gender-guesser}\footnote{https://pypi.org/project/gender-guesser/}. Name-based methods have been effectively applied to measure demographics in a variety of online contexts~\cite{mislove2011understanding}. Recent work using eye-tracking software suggests that individuals evaluating content online do look at names and photos of authors~\cite{ford2019beyond}.

Gender-guesser is a dictionary of over 40,000 first names (collected from a variety of countries and regions) and their most likely binary gender, sourced from public statistics of names and sex recorded at birth. 
Specifically, each name contained in the dictionary is described as \emph{male}, \emph{female}, \emph{mostly male} or \emph{mostly female}. We apply gender-guesser to the first token in each user name. We ignore the inference for the latter two categories, classifying reviewers with ``mostly male'' and ``mostly female'' names as having unknown gender. To widen the scope of our inference, we also consider a manually collected list of keywords that give a clear indication of a specific gender, but are not given names (such as  \emph{`girl'} or \emph{`woman'} for women and \emph{`boy'} or \emph{`dude'} for men).

Using this procedure, we classify from the $20.9$ million unique reviewers approximately $5.43$ million ($26.0\%$) as likely signaling men, and $5.6$ million ($26.8\%$) as likely signaling women. These account for around $18.9$ million and $19.3$ million reviews respectively. The remaining reviewers (ca. $11.03$ million or $47.2\%$, accounting for $41.1$ million reviews) do not signal a likely gender with their user name. 
As we are interested in how perceived gender of a reviewer relates to the social feedback given by other users, we believe that our approach, namely to infer gender from displayed name using frequency, gives an accurate picture about how other users perceive the gender of a reviewer.

\noindent\textbf{Comparison with human gender perception of user names. }
We confirm this assumption of user name categorization reflecting human gender perception in an experiment. 
For this purpose, we provided $6$ human annotators with $500$ user names that had been labeled with a gender by our procedure, and ask them to guess the genders. Typically, for a given user name annotators were given three choices - (i) `male', (ii) `female' and (iii) `can't say'. Further,  the responses of all the $6$ annotators were recorded for every user name.  
The annotators achieved an overall good inter-annotator agreement (`Fleiss'-kappa' of $0.81$) for this task. In case of disagreements between annotators, we assigned a label for human gender perception based on majority voting. As a result, we observe a match of 98\% ($490$ out of $500$) between our automatic gender inference and human gender labeling. Additionally, we asked human annotators to assign a gender to $500$ random user names, which we could not automatically classify. In this case, human annotators are unable to detect gender from names in $90$\% of the cases (`Fleiss'-kappa' of $0.8$). Inspection of cases where human annotators were able to guess the gender from these names show mostly names with unorthodox spelling but similar phonetics (e.g., 'Florentyna' instead of 'Florentina'). This suggests a future line of research to improve name-based gender inference tools.

Together, the above results indicate that our automatic procedure is well in line with human perception for detecting genders from user names, i.e., if we can infer a gender based on the user name automatically, the gender signal contained in a user name is strong, while there is no clear indication of the gender from the user name if we cannot assign a gender label automatically. In the remainder of the paper, we therefore differentiate between two groups of reviewers (and by implication, reviews):

\begin{itemize}
\item {\bf Signaling Men and Women}: This is the set of reviewers for which we can automatically infer the gender from the user name, i.e., the user name sends a clear signal about the likely gender of the reviewer.

\item{\bf Non-signaling users}: This is the set of reviews for which we cannot infer the gender from the user name. As shown by our experiments, humans do not pick up a clear gender signals for these users either. 
\end{itemize}

Note that we assign here a gender label based on the gender perception of the user name, i.e., we assume a review written under a name with a likely gender is a strong signal of gender identity. While there are exceptions, we believe this is a reasonable assumption that has been adopted previously in multiple studies ~\cite{liu2013s,ciot2013gender,mislove2011understanding}, see also Section \ref{sec:discussion}. Another point of validity for our assumption is that we are concerned with social feedback of other users in response to the signals contained in the user names. For the remainder of this paper, when we mention the signaled gender of a review, we would essentially mean the signaled \textit{likely} gender of the reviewer who authored the review unless specified otherwise.

\noindent
\subsection{Measuring gender performance}
\label{detect_gender}

\begin{figure*}[t]
\centering
\includegraphics[scale = 0.35]{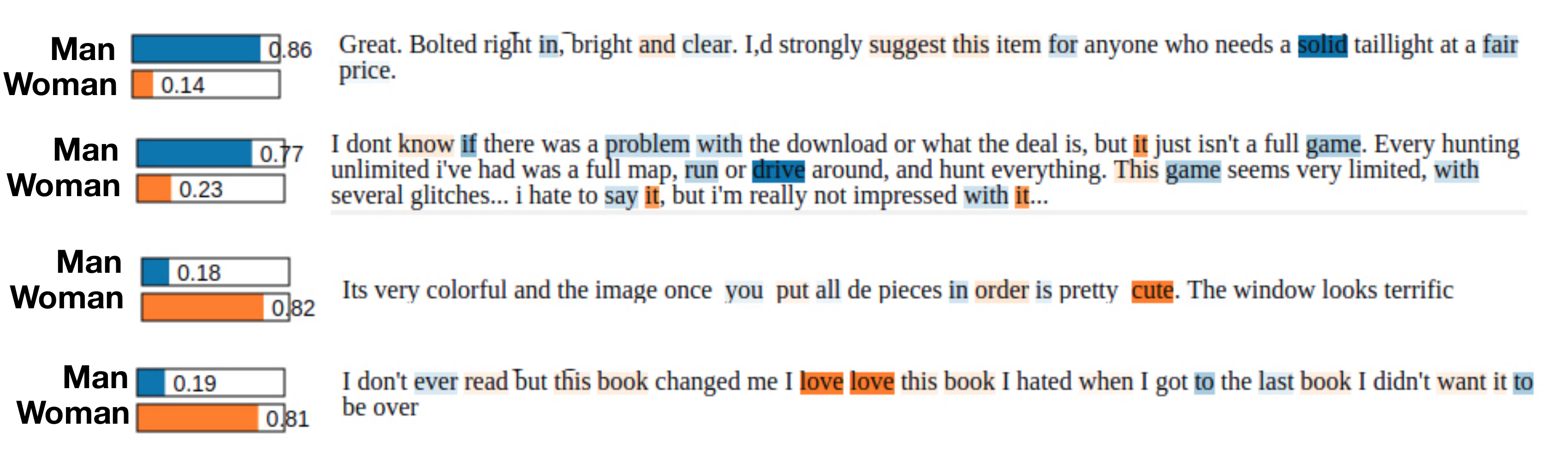}
\caption{\emph{Distinctive writing patterns of inferred men and women.}  The first two examples are cases where the LogReg model predicts that the author is a signaling man with high probabilities (0.86 and 0.77 respectively) while the classifier predicts that the second pair of texts were very likely written by signaling women. Noticeably, the model gives more weight to words such as `solid', 'drive' etc. for the first two cases where the model predicts the gender. For the next two examples the model gives more weight to words such as `love', `cute' etc. These exploratory results suggest that there exist differences in writing styles between gender signaling men and women while reviewing a product on Amazon.}
\label{text_exp}
\end{figure*}

%Since we intend to investigate the effect of disclosing gender, we need to obtain the genders of undisclosed reviews as well. Towards this goal,
In this work, we are interested in the effect of explicit gender signal on the perceived helpfulness.
Since differences in helpfulness scores between signaling men and women can originate from these direct signals or from different underlying gendered behavior (\textit{gender performance}), we next aim to model gender performance in review text.
For that purpose, the set of reviews with gender-signaling authors can be used as ground-truth to train machine learning models that infer the likely gender of the author from review texts. 
These would pick up on the gendered behavior embedded in the text of the review including style and word choice. However, designing such a classifier is only feasible if there is indeed a noticeable
difference in the writing patterns of men and women. For illustration, we first train a simple logistic regression classifier (LogReg) on the ground truth (gender-signaling) set of reviews. 
Then, by utilizing the LIME framework~\cite{ribeiro2016should} we compare the reviews which are inferred by the classifier as (authored by) men with those inferred as women. In Figure \ref{text_exp} we present four such examples (two in each class). For the reviews predicted as written by signaling men, the classifier gives higher weights to the words like `solid', `drive' and `game', while words such as `cute' and `love' are strong predictors that a review was written by a signaling woman. 
This goes to show that our collection of signaling men and women have distinctive writing styles which could be leveraged to design machine learning classifiers for inferring gendered behavior or performance from text.

In that direction, we now train a variety of machine learning models on the review text of reviews with disclosed gender (i.e., for which a gender is apparent from the user name) and apply the best performing model on the review text of reviews with undisclosed gender (i.e., for which the user name does not send a clear gender signal).
The underlying idea is that we can derive the gendered behavior of users in many cases with high accuracy using  machine learning techniques even if the underlying patterns used for classification cannot be picked up by humans. 

\noindent\textbf{Machine learning for inferring gender from the review text. }
In general, we aim to train a model on the review text of reviews with disclosed gender (i.e., for which we could identify a gender label based on the user name) and apply this model to infer a gender label for undisclosed reviews. For this text classification task, we considered a variety of traditional machine learning models such as Logistic Regression~\cite{fan2008liblinear} and Linear SVM~\cite{fan2008liblinear}, or XGBoost~\cite{friedman2001greedy} as well as state-of-the-art Deep Learning methods based on Recurrent Neural Networks (including LSTM~\cite{hochreiter1997long} and GRU~\cite{cho2014learning} or on Convolutional Neural Networks (CNNs) \cite{kim2014convolutional,zhang2015character}.
After extensive experiments, we focused on character-level CNNs since they offered the best predictive performance for our task by a small margin, and has also previously been shown to perform well for NLP tasks in general and classification tasks in particular. We describe the detailed setup of this model next.

The vocabulary for this model is elementary and consists of only 69 characters which include 26 English alphabets, 0-9 digits and 33 other special characters. As input, the text is quantized with the help of this vocabulary. For each review, we consider a maximum text length of 1014 characters (which is sufficient to cover most of the reviews), dropping exceeding characters or padding missing ones, cf. \cite{zhang2015character}. We train the model with a batch size of 512.
%Characters more than this are dropped and not considered. 
The model itself consists of overall 9 layers, i.e., 6 convolutional layers and 3 fully connected layers. As suggested by literature, the first two and sixth convolutional layers are followed by pooling layers to reduce the dimensionality. We use 2 dropout layers in between the fully connected layers for better regularization. The final predication is made by a sigmoid activation function and as loss function, we employ standard cross-entropy loss.

%\noindent{\bf Evaluation setup and results}: 
For evaluation, we split the labeled data into a training set of $\sim 28.7$ million or $80\%$  and a test set with the remaining reviews. The split was performed on a user level such that reviews from one user are all contained either in the training or in the test set. As a result, our classification model is able to correctly classify $75.2\%$ of all reviews. 
As a subsequent step, for each user we aggregate the predicted gender label of all her (his) authored reviews (recall that a gender label is assigned to each review independently) and then assign her (him) a gender behavior score through majority voting. Once a user is labelled we assign each review the gender of its corresponding author (user). This enhances the gender prediction accuracy to $82.2\%$.

\noindent\textbf{Comparison with human gender perception of review text. }
In this work, we assume that in general humans do not directly identify a reviewer's gender from her/his review texts, even though advanced machine learning algorithms are capable of nontrivial accuracy. To check this assumption, we randomly sampled 300 (132 performing men and 168 performing women) reviews with disclosed gender from our dataset and asked three volunteers to infer the gender of the author of each review based on its text only. Volunteers were provided four options for their responses -- `definitely male', `definitely female', `probably male' and `probably female'. The inference is made using a majority voting. 

We present the result of the experiment in Table~\ref{tab_he}. For 55 cases (which accounts for 18.3\% of the total set), agreement could not be reached among the participants regarding the gender of the author. For only 18 cases (accounting for 6\%), the participants could judge the author to be `definitely male'. However, 17 of them were indeed signaling men from their username. 
Gender was judged to be `probably male' in 75 cases (25\%) of which 64 were performing men and 11 were performing women. For 35 (11.7\%) cases, the participants assigned the gender of the author to be women and were correct in all cases. Finally 117 (39\%) cases were inferred as `probably female' of which 88 were indeed women while 29 were men. 
%The results demonstrate that humans are better at identifying performing women authors than performing men. 
Moreover, all these cases were predicted correctly with high probability by our model. This indicates that machines are better at predicting gender from text than humans. 

\begin{table}[]
\centering
    \small
    \setlength\tabcolsep{1px}
\begin{tabular}{c||c|c|c|c|c||c}
\hline
       & \rotatebox{90}{No Majority}                                           & \rotatebox{90}{Definitely Male}                                       & \rotatebox{90}{Probably Male}                                         & \rotatebox{90}{Definitely Female}                                     & \rotatebox{90}{Probably Female}                                       & Total \\ \hline
\textcolor{blue}{Male}   & \begin{tabular}[c]{@{}c@{}}22\\ (16.7\%)\end{tabular} & \begin{tabular}[c]{@{}c@{}}17\\ (12.9\%)\end{tabular} & \begin{tabular}[c]{@{}c@{}}64\\ (48.5\%)\end{tabular} & \begin{tabular}[c]{@{}c@{}}0\\ (0.0\%)\end{tabular}   & \begin{tabular}[c]{@{}c@{}}29\\ (21.9\%)\end{tabular}   & \textcolor{blue}{132}   \\ \hline
\textcolor{red}{Female} & \begin{tabular}[c]{@{}c@{}}33\\ (19.6\%)\end{tabular} & \begin{tabular}[c]{@{}c@{}}1\\ (0.6\%)\end{tabular}   & \begin{tabular}[c]{@{}c@{}}11\\ (6.5\%)\end{tabular}  & \begin{tabular}[c]{@{}c@{}}35\\ (20.9\%)\end{tabular} & \begin{tabular}[c]{@{}c@{}}88\\ (52.4\%)\end{tabular} & \textcolor{red}{168}   \\ \hline
Total  & \begin{tabular}[c]{@{}c@{}}55\\ (18.3\%)\end{tabular} & \begin{tabular}[c]{@{}c@{}}18\\ (6\%)\end{tabular}    & \begin{tabular}[c]{@{}c@{}}75\\ (25\%)\end{tabular}   & \begin{tabular}[c]{@{}c@{}}35\\ (11.7\%)\end{tabular} & \begin{tabular}[c]{@{}c@{}}117\\ (39\%)\end{tabular}  & 300   \\ \hline
\end{tabular}
\caption{\label{tab_he} Results from the survey experiment indicate that humans are indeed not very good at predicting gender from text.}
\end{table}

\noindent

\noindent\textbf{Categorization of reviewers}
We now leverage the trained model to infer the performed gender of the authors of the reviews in the undisclosed set (i.e., for which the user name does not send a clear gender signal). 
Note that we only consider the reviews where our model is able to infer gender with probability of at least $0.7$ (reduces the undisclosed set to $\sim$ 29 million). Further, for a given user we aggregate gender information across all her(his) reviews and assign gender through majority voting.  
The whole set of reviews now can be divided into four categories - 

\begin{itemize}
\item {\bf Signaling (likely) man}: Reviews authored by users for which we can infer that they are likely men from their user name.
\item {\bf Signaling (likely) woman}: Reviews authored by users for which we can infer that they are likely women from their user name.
\item {\bf Performing (likely) man}: Reviews authored by users who do not signal gender with their user name and for which the text-based classifier identifies them as men with probability of at least 0.7.
\item {\bf Performing (likely) woman}: Reviews authored by users who do not signal gender with their user name and for which the text-based classifier identifies them as men with probability of at least 0.7.
\end{itemize}

For the sake of readability, we drop ``likely'' labels from the four categories.

\begin{figure*}
\begin{center}
\subfigure[Upvotes]{\label{fig:upvote}\includegraphics[scale=0.14]{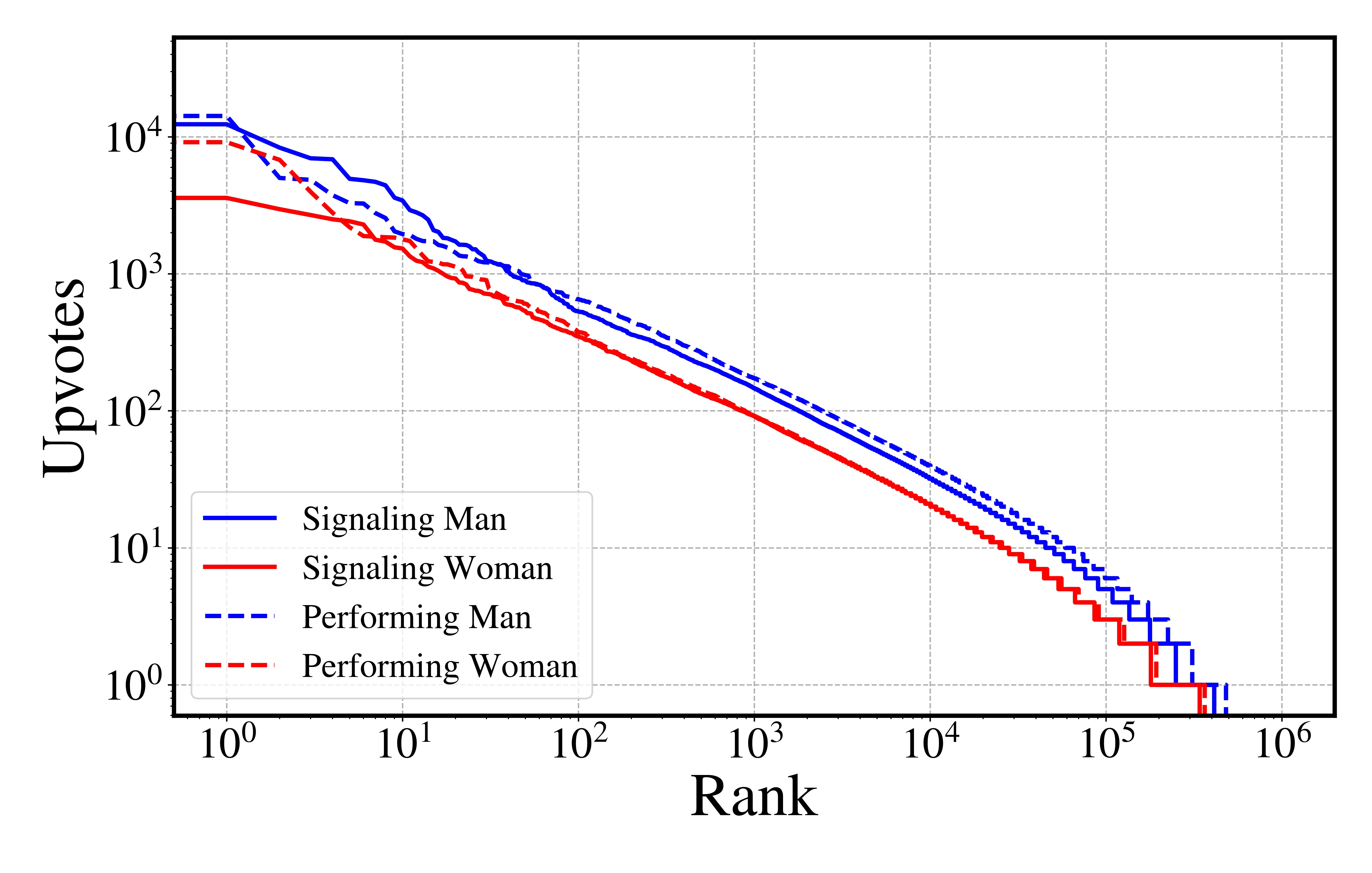}}
\subfigure[Downvotes]{\label{fig:downvote}\includegraphics[scale=0.14]{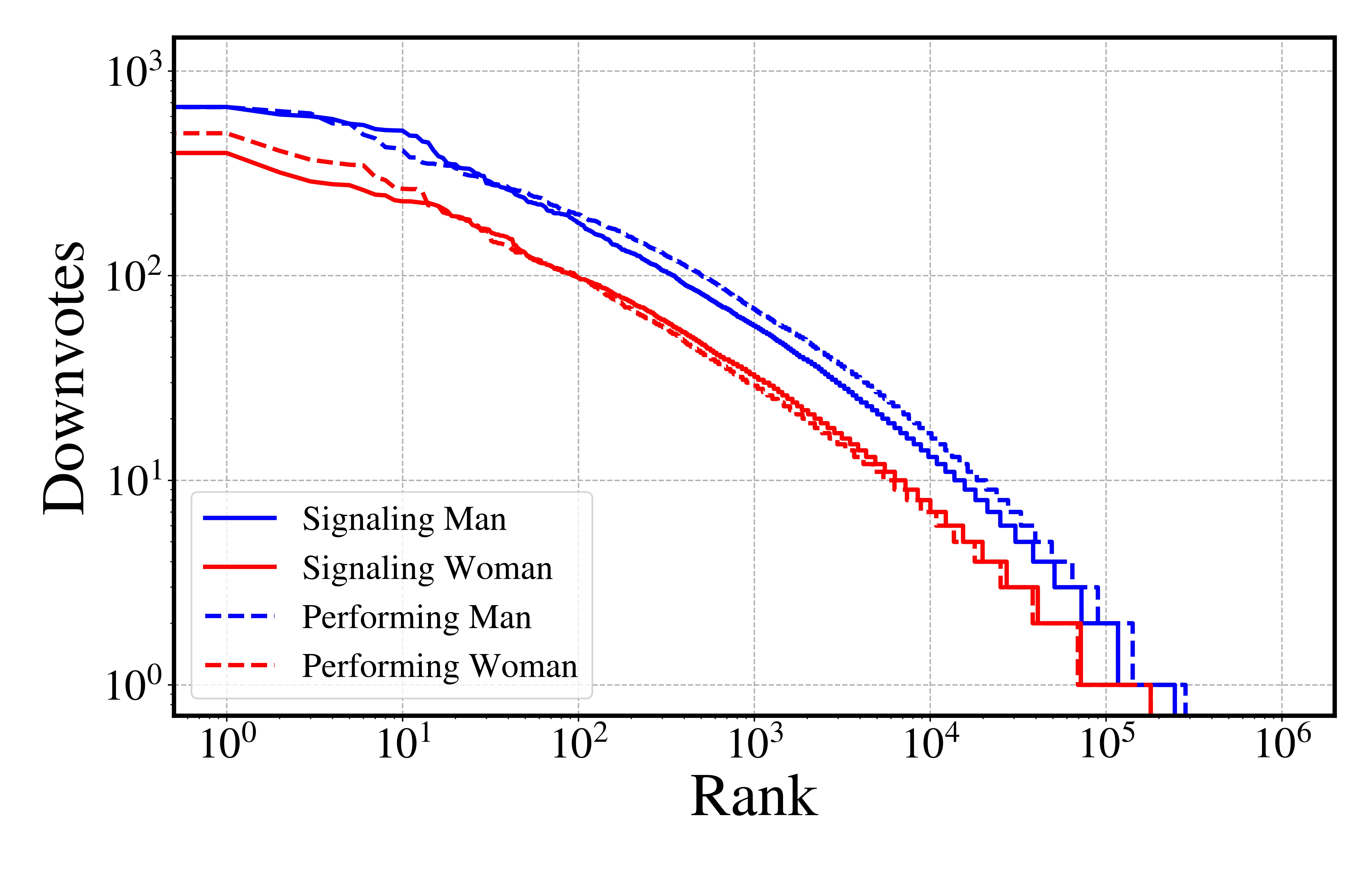}}
\subfigure[Helpfulness]{\label{fig:helpfulness}\includegraphics[scale=0.14]{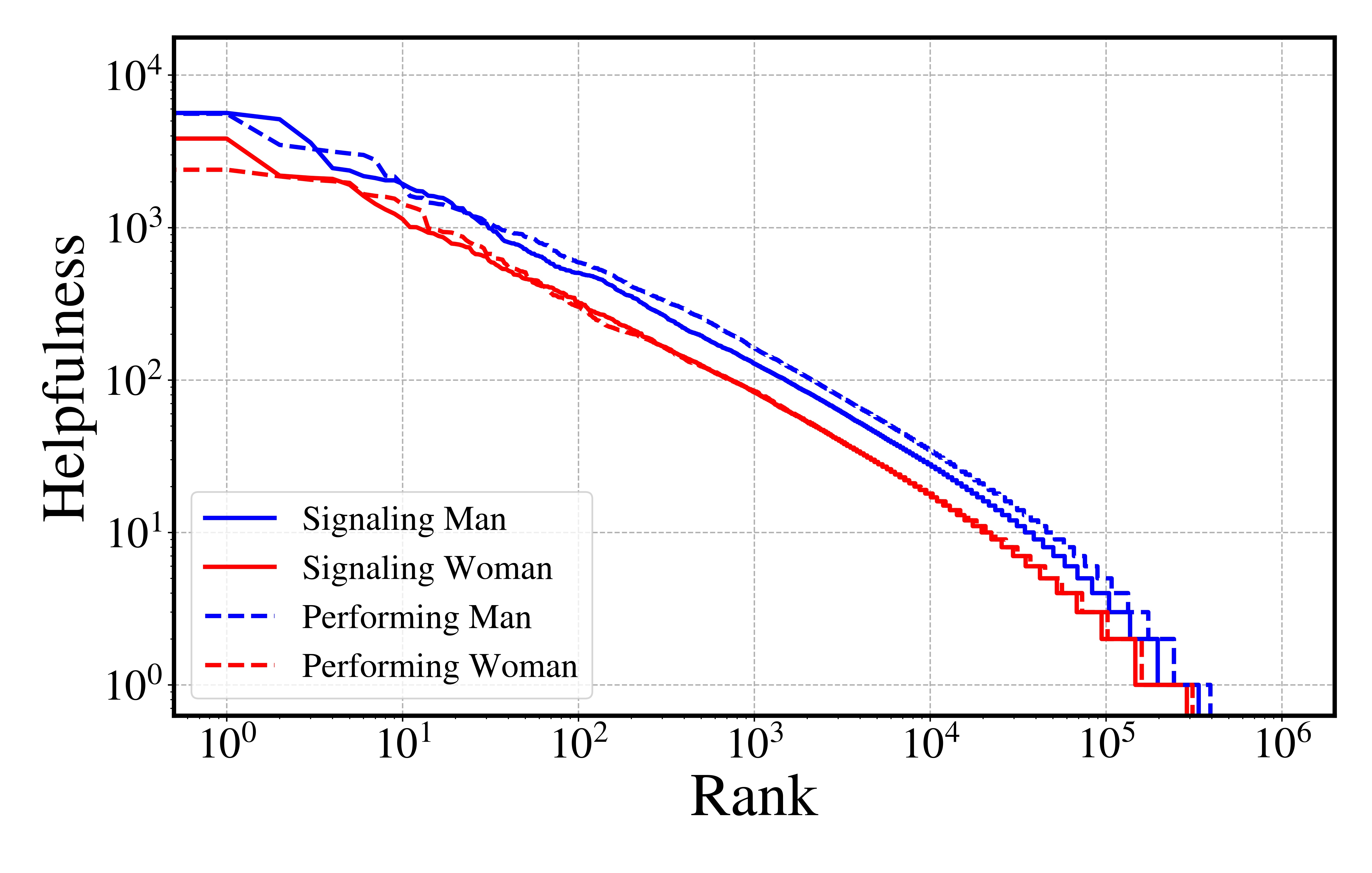}}

\end{center}
\caption{\emph{Comparing perceived helpfulness} We sample $1$ million reviews from each of the four categories and rank them based on upvotes, downvotes and helpfulness score (\#upvotes - \#downvotes). We plot the rank and the corresponding values of (a) upvotes, (b) downvotes and (c) helpfulness scores for the sampled reviews. On average reviews authored by signaling men tend to receive more upvotes as well as more downvotes. They are also perceived as more helpful on average.}
\end{figure*}

\subsection{Gender differences in perceived helpfulness}
We now consider the differences in perceived helpfulness between reviews authored by likely men and women as well as differences between users signaling gender versus those performing gender without signaling. To this end, we first sample $1$ million reviews from each of the four sets ( \emph{signaling men}, \emph{signaling women}, \emph{performing men}, and \emph{performing women}) and then rank them based on upvotes, downvotes and helpfulness score (\#upvotes - \#downvotes). In figures \ref{fig:upvote},~\ref{fig:downvote} and \ref{fig:helpfulness} we plot the rank and the corresponding value for the metrics upvotes, downvotes and helpfulness respectively. We observe that reviews authored by signaling men tend to receive higher upvotes as well as downvotes irrespective of whether gender information is available. Similar observations are made for helpfulness score as well. 

\medskip

\section{Method}
To determine adjust for potential confounders in the relationship between gender signaling and perceived helpfulness, we employ matching experiments which we describe next.

\noindent\textbf{Treatment groups. } Each review in the dataset is classified into one of the four groups -  \emph{signaling men}, \emph{signaling women}, \emph{performing men}, and \emph{performing women}.

\noindent\textbf{Matching reviews. } Ideally, to eliminate potential confounders we would like to compare the perceived helpfulness of a review when authored by an individual of one group (e.g., signaling man) with a review with the same properties when authored by an individual from the other group (e.g., signaling woman). Since a review can only be authored by an individual belonging to exactly one group and it is also unlikely for two reviews to be exactly same, we manually identify a set of potential confounder variables (i.e., variable whose presence affects the outcome of the variable being studied) and control for them. For that purpose, we leverage Mahalanobis matching on a set of confounders (factors that directly influences the perceived helpfulness of a review) for each review to obtain similar review pairs. 

When comparing groups $S_1$ and $S_2$, we randomly select a review authored by an individual in $S_1$ and obtain the most similar review from $S_2$. Typically, the similarity is measured in terms of Mahalanobis distance on the following confounders - (1) time when the review was published, (2) length (in terms of number of words), (3) readability, (4) sentiment and (5) overall rating. While (1) ensures that both reviews on the matched reviews had approximately equal time of exposure (2), (3) and (4) ensure they are of similar quality. We further ensure that the reviews were written on same product category (6). Note that we use Mahalanobis distance matching as it allows for variable standardization by including sample covariance matrix in distance calculation. We considered using propensity score matching but rejected this approach given the recent problems highlighted by \cite{king2019propensity}. In fact, the authors demonstrate that propensity score matching can increase imbalance, model dependence, and bias.
%\fle{Describe sampling, mention that Mahal. distance includes variable standardization?}

\noindent\textbf{Paired-treatment groups. } We consider four paired-treatment groups, comparing how helpfulness is perceived when:
\begin{enumerate}
    \item \textbf{PM-PW} a review is authored by a performing man vis-a-vis when authored by a performing woman and user names do not signal gender.
    \item \textbf{SW-SM} a review when authored by a signaling woman vis-a-vis when authored by a signaling man, both inferred from user name.
    \item \textbf{SW-PW} a review authored by a signaling woman vis-a-vis when authored by a performing woman that does not signal gender.
    \item \textbf{SM-PM}  a review authored by a signaling man vis-a-vis when authored by a performing man that does not signal gender.
\end{enumerate}
The first case compares the effects of performed gendered behavior on differences in outcomes. The second compares the effects of signaled gender on differences in outcomes. The remaining two cases measure the advantage (resp. penalty) gained (resp. paid) for signaling gender information.

\begin{figure}[!t]
  \begin{center}
    \subfigure[Readability]{\label{fig:read_co}\includegraphics[width=0.4\columnwidth]{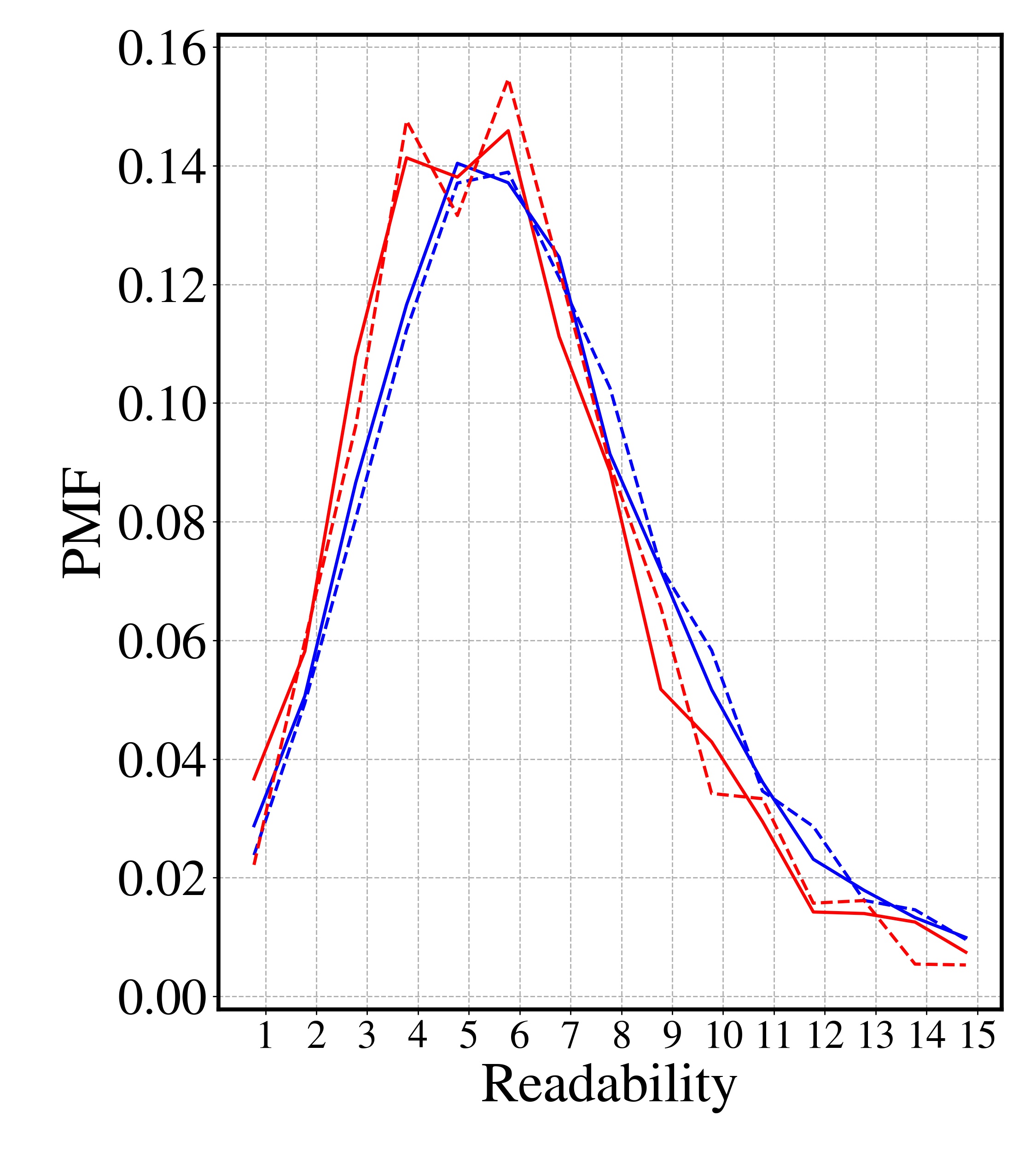}} 
    \subfigure[Rating]{\label{fig:rat_co}\includegraphics[width=0.4\columnwidth]{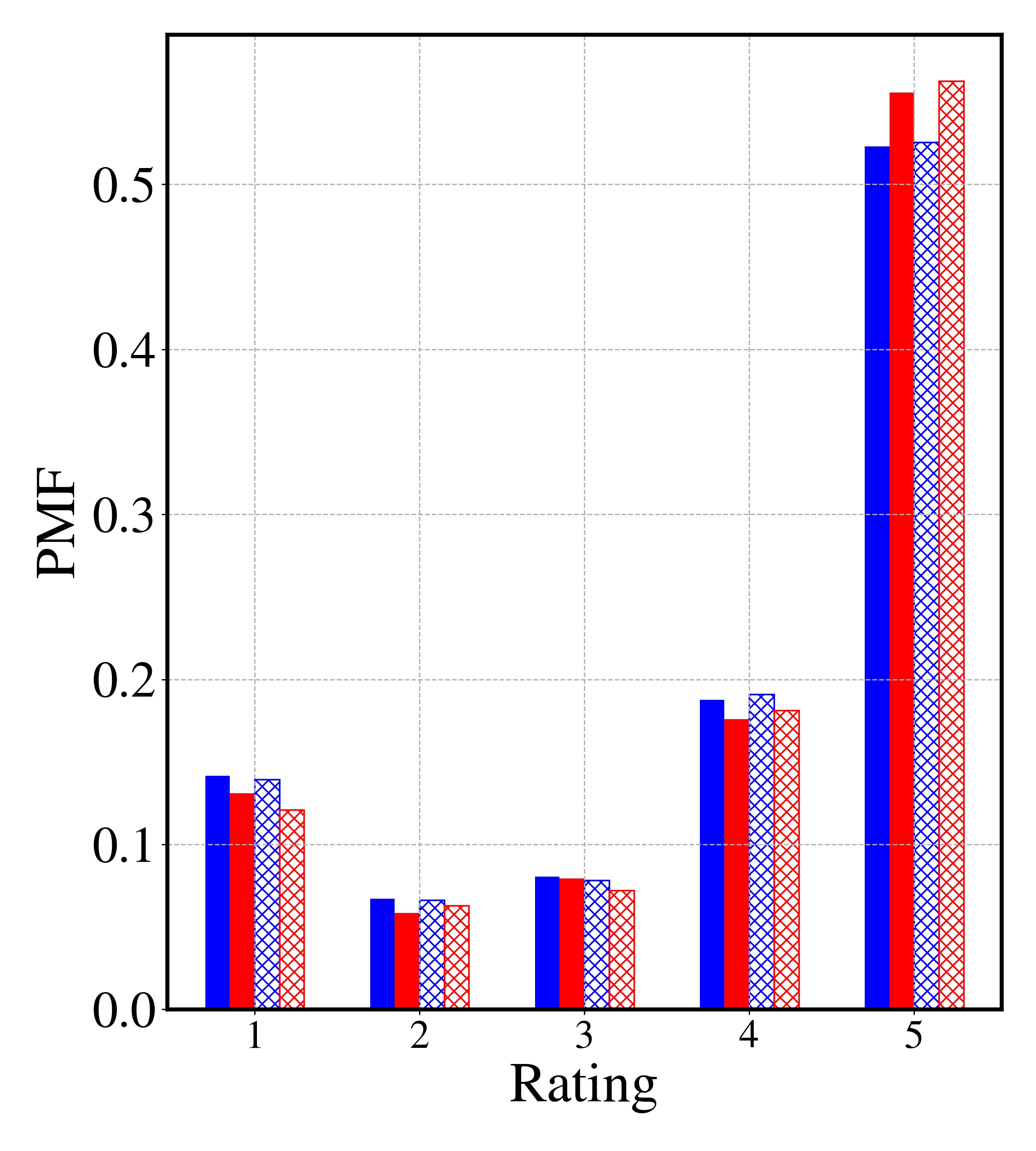}}
  \end{center}
  \caption{\emph{Distribution of the matching confounders.} We present the distributions of two exemplary potential confounders (a) Readability score and (b) overall rating for the reviews in the matched set across the four groups for the category {\myf Books}. Note that for the two confounder (results are similar for others) the distribution across the for groups closely resemble each other. This shows that the reviews in the matched set are indeed similar and hence inference drawn on these matched sets are consistent.}
  \label{fig:con_dist}
\end{figure}

\begin{figure*}[t]
  \begin{center}
    \subfigure[PW-PM]{\label{fig:PW_PM}\includegraphics[scale=0.12]{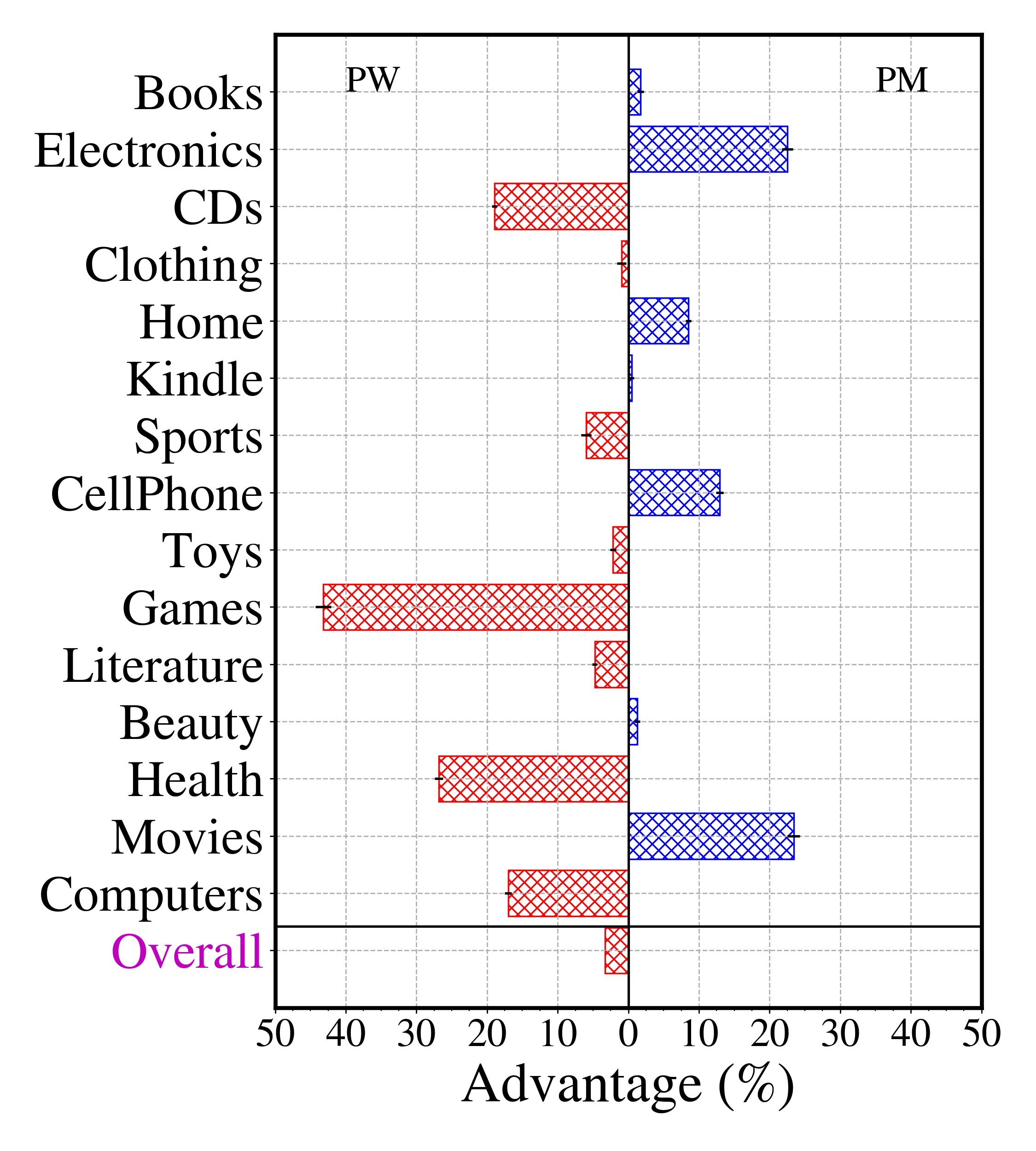}}
    \subfigure[SW-SM]{\label{fig:SW_SM}\includegraphics[scale=0.12]{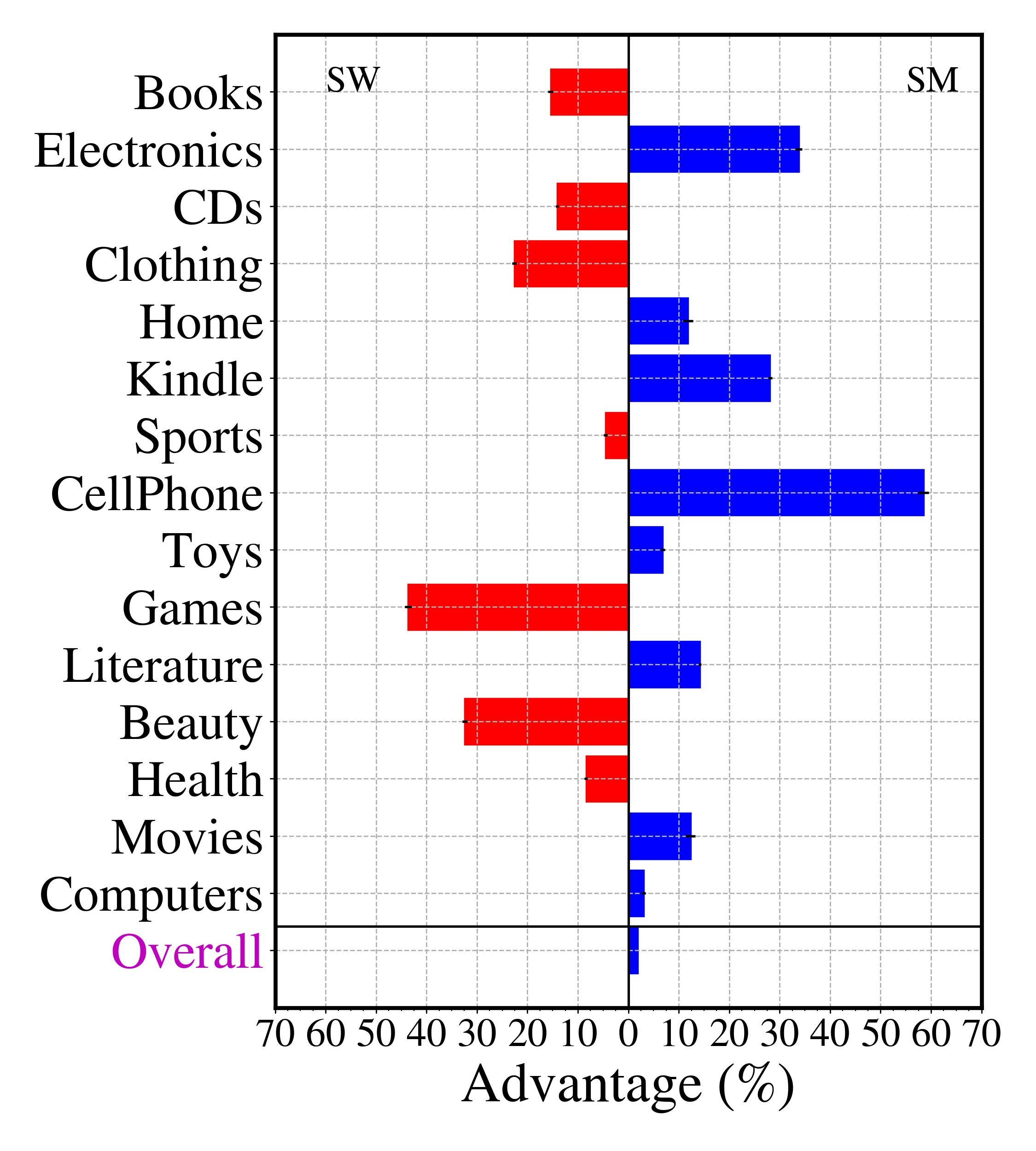}} 
    \subfigure[PW-SW]{\label{fig:PW_SW}\includegraphics[scale=0.12]{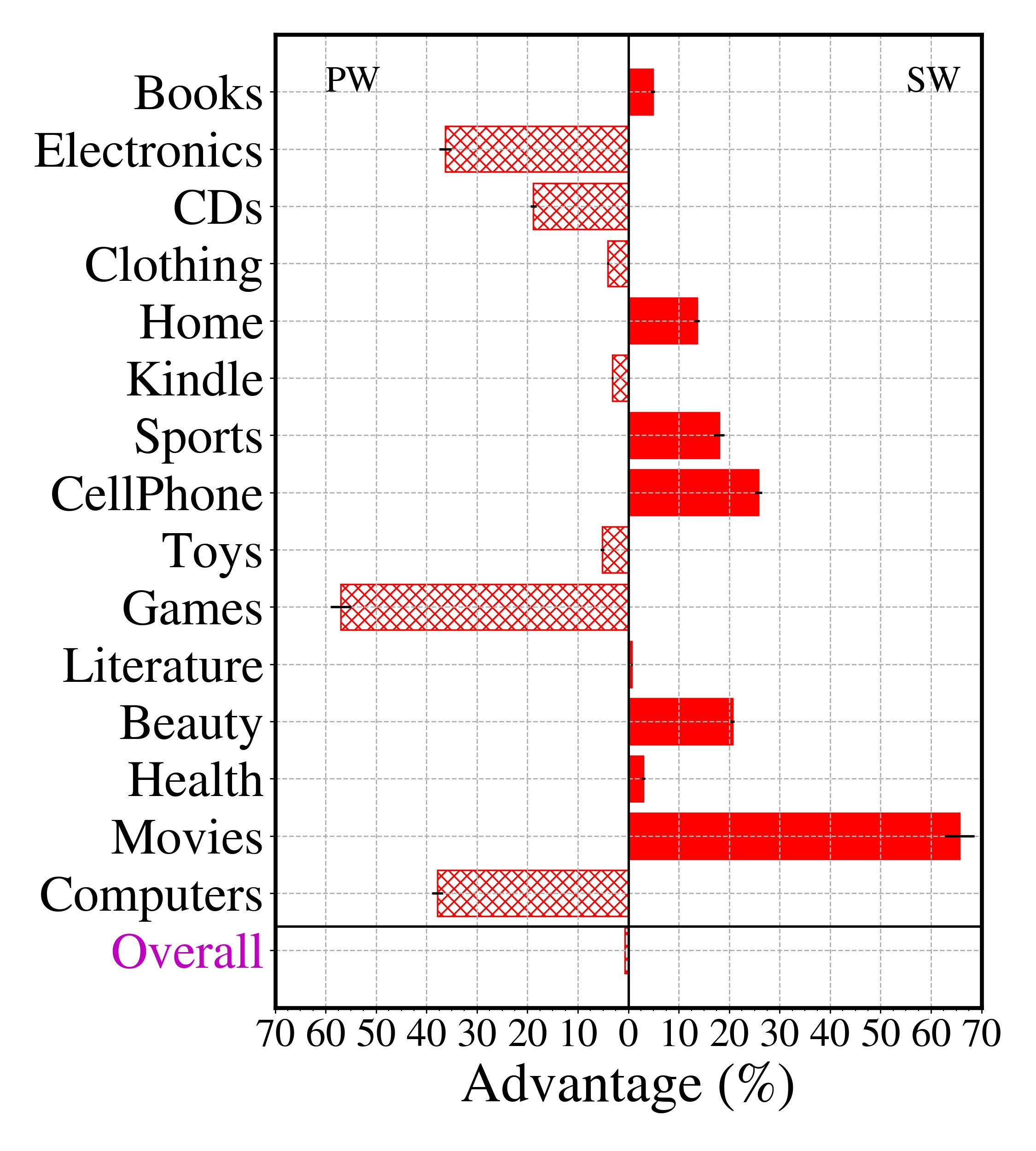}}
    \subfigure[PM-SM]{\label{fig:PM_SM}\includegraphics[scale=0.12]{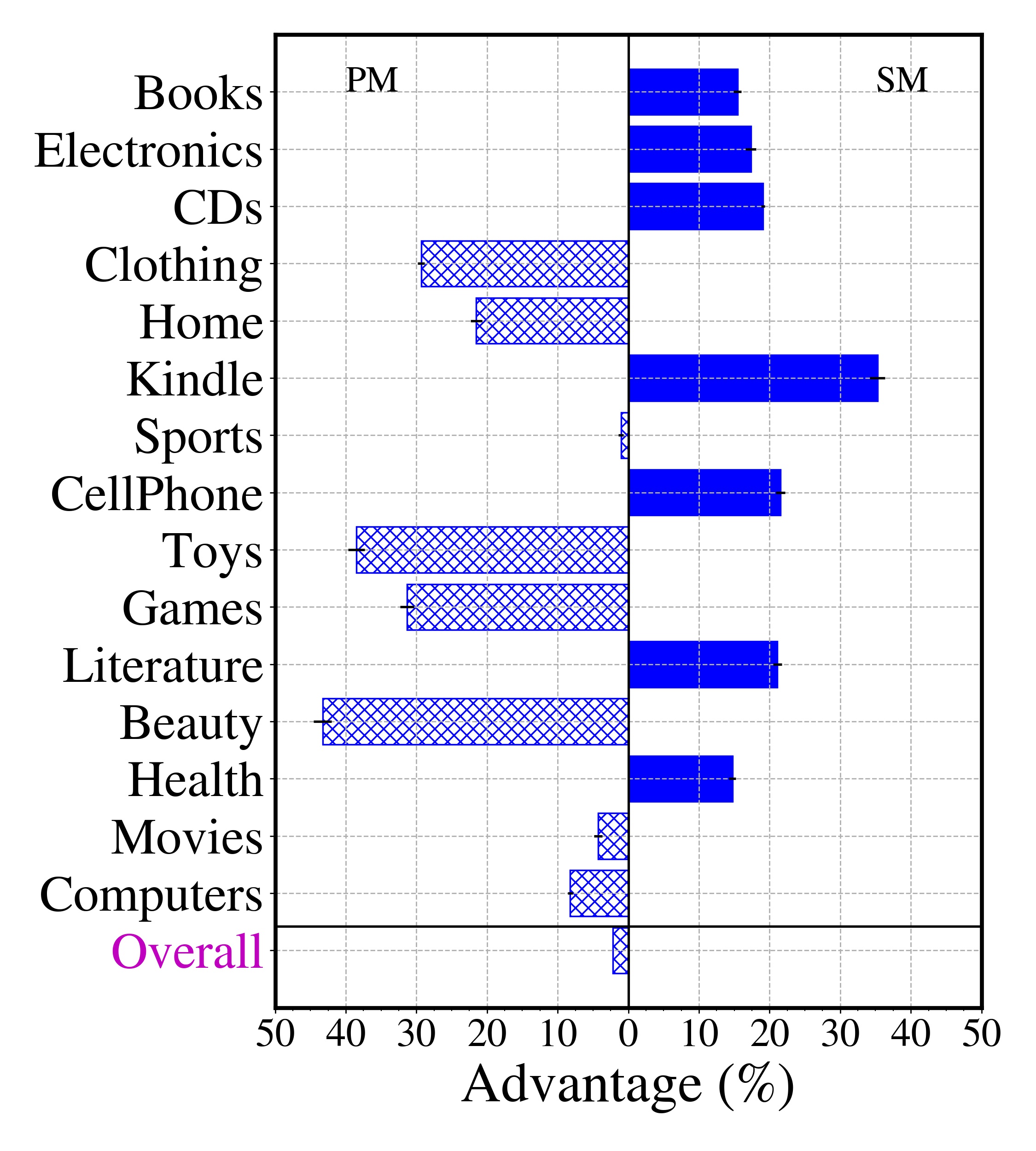}}
  \end{center}
  \caption{\emph{The effects of gender signaling and performance.} We show the advantage gained by one group over the other across different categories when comparing the paired-treatment groups - (a) performing women and performing men (PW-PM), (b) signaling women, signaling men (SW-SM), (c) performing women, signaling women (PW-SW) and (d) performing men, signaling men (PM-SM). Notably reviewers with unclear gender signals in their user name but performing as women receive higher scores than signaling women in categories such as {\myf Electronics}, {\myf Games} or {\myf Computers} while the same is true for reviewers performing as men with unclear gender signals in categories like {\myf Clothing}, {\myf Beauty} or {\myf Toys}. However, \emph{overall}, there seems to be no effect of gender signaling on the perceived helpfulness of a review in general. Note that in each case the advantage reported is calculated as mean over bootstrap samples. We also report the standard error of the means which are very low for all the cases across four paired-treatment groups}
  \label{fig:nat_expt}
\end{figure*}

\noindent\textbf{Balancing paired-treatment groups. }
First we randomly sample $N=10,000$ instances from the union of all the treatment groups. We use sampling since the following matching procedure is computationally unfeasible for the complete dataset. Moreover, the number of reviews in each category is still large, and hence  results on a random sample are  still representative for the overall population.

Note that by doing so, the initial random sample is not balanced between the four groups but reflects the overall distribution in the data. Then, we identify for each review the most similar review from each of the other groups across the complete set of reviews, i.e. also outside of the initial sample. So for a group $S_1$ (e.g., signaling man), we obtain matching reviews from others groups $S_2$ and $S_3$ (e.g., signaling woman and performing woman). We obtain matched review sets for the four paired-treatment groups.

\noindent\textbf{Comparing outcomes.} Note that each pair of reviews obtained for a paired-treatment groups (obtained in previous  step) are analogous to each other in terms of review quality and times of exposure and should ideally elicit similar perception among users. For a set of $N$ matched reviews for a given paired-treatment group $\{S_1 \times S_2\}$, we calculate the mean helpfulness score ($|$upvotes$|$ - $|$downvotes$|$) for the reviews in $S_1$ ($h_{S_1}$) and $S_2$ ($h_{S_2}$). The advantage of one group over the other is then denoted as $\frac{|h_{S_2} - h_{S_1}|}{min(h_{S_2},h_{S_1})}\times 100$. If the gender of the author does not influence how a review is perceived, the advantage should be negligible. To investigate the robustness of the results, we perform bootstrap sampling on the sampled pairs.  Typically, for each paired-treatment group, once we have sampled $N$ pairs, we perform a sampling with replacement from this $N$ sampled pairs. This leads to a resampled set of the $N$ pairs. The procedure is then repeated $N$ times to obtain $N$ bootstrap samples each of size $N$. We report the mean advantage and the standard error calculated over these bootstrap samples.  

\section{Results}

As mentioned previously, we consider four paired-treatment groups. The experiments are carried out across 15 different categories  (selected based on the number of reviews as well as diversity) - (1) {\myf Books}, (2) {\myf Electronics}, (3) {\myf CDs}, (4) {\myf Clothing}, (5) {\myf Home}, (6) {\myf Kindle}, (7) {\myf Sports}, (8) {\myf Cellphone}, (9) {\myf Toys}, (10) {\myf Games}, (11) {\myf Literature}, (12) {\myf Beauty}, (13) {\myf Health}, (14) {\myf Movies} and (15) {\myf Computers}. We now look into each paired-treatment group in detail. 

To start with, we look into the distribution of confounders in the matched sets across the four groups. In figure~\ref{fig:con_dist}, we plot the distribution of readability and ratings of the reviews  written on {\myf Books} across the four groups. We observe, the distributions closely resemble each other across all the four groups. The results are similar for other confounders as well. This demonstrates the reviews in the matched set are indeed similar and the inferences drawn from the matching experiments are consistent.  

In Figures \ref{fig:PW_PM},~\ref{fig:SW_SM},~\ref{fig:PW_SW} and \ref{fig:PM_SM} we plot the advantage (mean calculated over the bootstrap samples) of one group over the other. We also report the standard error for the average advantages in the same figures. The errors are low in almost all cases. Overall, across all categories we do not find any critical advantage of one group over the other which means gender signaling does not have a compelling effect on the perceived helpfulness of a review. However, we do observe within-category effects.

\noindent\textbf{Performing women - performing men (PW-PM)}: We plot the advantage of one group over the other in figure \ref{fig:PW_PM}. Performing men reviewers have advantage over their women counterparts when writing reviews for products on {\myf Movies} ($23.8$), {\myf Electronics} ($20.7$), {\myf Cellphone} ($14.8$) and {\myf Home} ($9.81$) while the opposite is observed in case of {\myf Games} ($44.2$), {\myf Health} ($27.6$)  {\myf CDs} ($18.8$) and {\myf Computers} ($17.4$). For the other categories the advantage is of one group over the other is marginal if any. Overall, across all categories there is no significant advantage for one group over the other.

\noindent\textbf{Signaling women - signaling men (SW-SM)}: We next compare the cases where reviewers signal gender via their user names. Although there does not seem to be any advantage overall for any particular group, 
we do observe significant advantages across individual categories (refer to figure \ref{fig:SW_SM}). Likely women get higher helpfulness scores in {\myf Game} ($44.9$), {\myf Beauty} ($32.1$) and {\myf Clothing} ($22.5$). Signaling men gain advantages over signaling women in categories {\myf Cellphone} ($60.5$), {\myf Electronics} ($35.6$), {\myf Kindle} ($28.5$) and {\myf Movies} ($12.1$). For other categories the advantages are marginal.

\noindent\textbf{Performing women - signaling women (PW-SW)}: We now consider the paired-treatment group PW-SW which allows us to probe into whether signaling women who do better than performing women reviewers. In figure \ref{fig:PW_SW} we plot the advantage gained by one group over the other across different categories. We note that performing women reviewers have better outcomes than signaling women in categories like {\myf Electronics} ($38.5$), {\myf Games} ($56.8$) and {\myf Computers}  ($38.8$). For categories like {\myf Movies} ($62.2$), {\myf Cellphone} ($25.2$) and {\myf Beauty} ($19.8$) the opposite effect is observed. For other categories the advantage is marginal if any. However, there is no advantage for any group on average overall. 

\noindent\textbf{Performing men - signaling men (PM-SM)}: Finally, we investigate differences in helpfulness scores between signaling men and performing men reviews. We observe (refer to figure \ref{fig:PM_SM}) that signaling men do better than performing men reviewers in categories like {\myf Kindle} ($33.6$), {\myf Cellphone} ($21.6$), {\myf CDs} ($17.5$) and {\myf Electronics} ($16.4$). In categories like {\myf Beauty} ($42.4$), {\myf Toys} ($35.3$), {\myf Clothing} ($29.1$) and {\myf Games} ($31.2$) the opposite is true. Again, we observe only negligible advantage overall on average.

\noindent
\section{Discussion}
\label{sec:discussion}
In this section we present the implications as well as few limitations of our study. 
%We also point toward certain future directions. 
%\noindent\textbf{Perceived vs. real gender}.
%\fle{We are missing the discussion of results and limitations right now completely!}
\subsection{Implications of results}
For categories like {\myf Electronics}, we observe that the signaling women with get less positive feedback than similar signaling men. They also get less positive feedback than performing women. This suggests a disclosure penalty, i.e. that other users consider the signal encoded user names when judging review helpfulness. Similarly, for categories such as {\myf Beauty}, signaling men are at disadvantage. One potential mechanism for this effect is that users judging reviews apply gender stereotypes -- for example that men are more knowledgeable about electronics while women are better informed about beauty products -- to rate reviewers when they can infer gender from names. As these ratings influence the ranking and visibility of reviews to shoppers, this can amplify stereotypes.

\begin{figure}[htpb]
    \centering
    \includegraphics[scale=0.52]{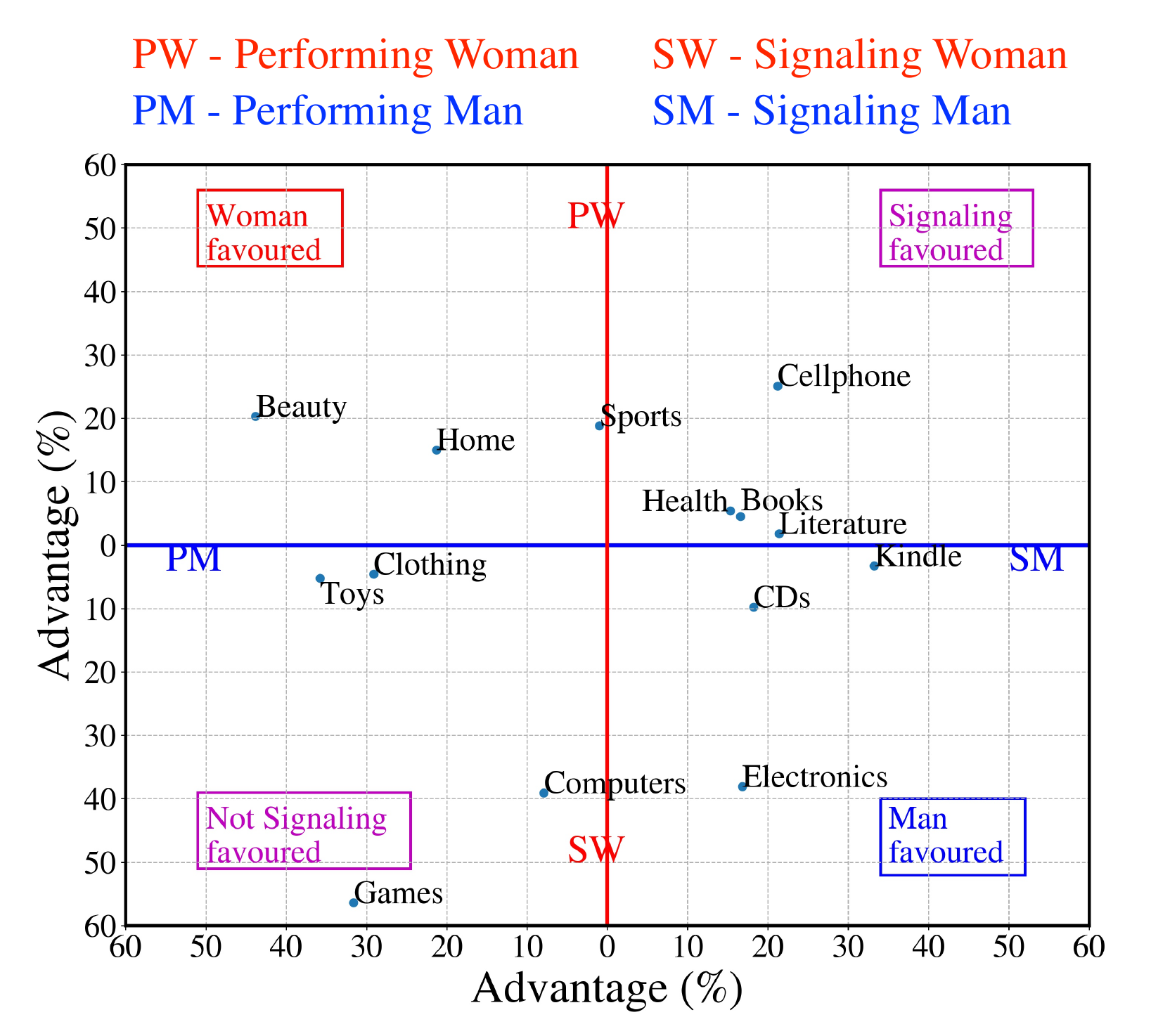}
    \caption{\emph{Summary of the context-specific gender effects.} We combine and summarize the results of figures \ref{fig:PW_SW} and \ref{fig:PM_SM} to classify each category into - (i) signaling man favoured, (ii) signaling woman favoured, (iii) signaling favoured and (iv) performance favoured. The x-axis from left to right denote performing to signaling men (PM - SM) while y-axis from top to bottom denotes performing to signaling women (PW - SW). Each point in the figure denotes a category and its position is determined by its corresponding value in figures \ref{fig:PW_SW} and \ref{fig:PM_SM}. Notably, categories like {\myf Electronics}, {\myf CDs} or {\myf Kindle} seem to favour signaling men. Similarly {\myf Beauty} or {\myf Home} seem to favor signaling women. Moreover, categories such as {\myf Cellphone}, {\myf Health} or {\myf Books} favour gender signals from user names while {\myf Games}, {\myf Computers} and {\myf Toy} favour non-signaling.} 
    \label{fig:summary}
\end{figure}

There is no global advantage or disadvantage for those users whose avatars do not signal a gender. However, such effects are observed within individual categories. We classify each category into one of four groups - (i) signaling man favoured, (ii) signaling woman favoured, (iii) gender-signalled favoured and (iv) non-disclosure favoured based on their advantage score for (un)disclosing gender information (refer to figures \ref{fig:PW_SW} and \ref{fig:PM_SM}). For example in {\myf Electronics} category signaling men hold an advantage of $16.4$ over performing men while performing women secure an advantage of $38.4$ over their signaling women counterparts. This places the {\myf Electronics} in the quadrant man favoured with coordinates ($16.4$, $38.5$). 
We make a few observations:
\begin{itemize}
    \item Categories like {\myf Beauty} and {\myf Home} seem to favour women as gender signaling via names increases helpfulness compared with performed gender while performing men do better than their signaling counterparts. The exact opposite holds for categories such as {\myf Electronics}, {\myf CDs} and {\myf kindle} (refer to figure \ref{fig:summary}).
    
    \item Categories like {\myf Cellphone}, {\myf Health} and {\myf Books} seem to favour users signaling gender via their names. Similarly categories such as {\myf Computers}, {\myf Games} and {\myf Clothing} seem to favour gendered performance without signals from user names (refer to figure \ref{fig:summary}).
\end{itemize}

%\fle{What do our results mean? What are the implications?}
\subsection{Limitations}
\label{sec:limitations}
%\fle{I just brainstormed a couple of points for discussion}
%\noindent\textbf{Real vs. perceived gender.} We generally consider that the users tend to use names which signal their true gender i.e., (fe)males (real gender) mostly use (fe)male-sounding names (perceived gender).  
%In fact, our results on entirety is based on this assumption. The best way of validating this hypothesis would be compare 
\noindent\textbf{User names and real gender.} We generally consider that the user names signal the true gender of the user. This assumption has been adopted previously in multiple studies and has shown practical relevance, despite various shortcomings~\cite{keyes2018misgendering}. A second simplifying assumption we make is that gender is binary. As we are measuring the social feedback received by reviewers, the salient gender feature is rather how a user's gender is perceived, rather than how the user identifies. A greater matter of concern is the known western bias of name-based gender inference tools including gender-guesser~\cite{qiu2019going}. As our data comes from a platform based in North America, we argue that this limitation is acceptable, but we certainly acknowledge that extension of our analyses to other regions will require careful modification of the gender inference approach.

\noindent\textbf{Disclosure of gender.}
One important point to consider is whether usernames are considered at all while assessing the helpfulness of a review. Past work using surveys and eye-tracking software indicates that users do notice and reflect on social signals when evaluating online content~\cite{marlow2013impression,ford2019beyond}. Although our results suggest there is indeed a dependency between gender signaling and perceived helpfulness, further analysis is required in this regard. In fact, our results may suggest that some users face incentives to hide or conceal their gender~\cite{vasilescu2012gender}. 

\noindent\textbf{Inferred gender behavior through machine learning.} For the undisclosed set we utilize our trained machine learning model to infer gendered behavior from the text of the reviews. Although our model seems to perform well (accuracy of $82.2\%$) on the ground truth (disclosed set), we cannot assess its performance on the undisclosed set. Likely there is a natural limit of the extent to which behavior conforms to gender identity. This limit is probably highly dependent on socio-cultural norms and so is changing all the time. Implicit in our approach is the assumption that gender signals from usernames and gender performance measured from content align or overlap. Future work is needed to better understand the complex interaction between presentation and performance of gender and its effects on online feedback.

\noindent\textbf{The mechanisms behind observed differences.} Our experiments can not reveal why disclosing or performing gender relates to different outcomes in different characters. While our findings do suggest that sometimes signaling gender does relate with better outcomes, it is unclear if this is because of audience demographics and preferences or bias. More work is needed to understand the process by which individuals rate reviews.

\noindent\textbf{Experimental limitations.} In this work we have used Mahalanobis Distance Matching as the preferred matching method. However, to test the robustness of the method one needs to look into \emph{additional matching methods} like coarse exact matching or caliper-based approaches. Moreover, we consider a set of six confounders of which readability, sentiment and length are selected to determine the quality of the review. However, more specific linguistic and psychological (presence of `insightful', `causation' or  `inclusive' words) dimensions could be used as \emph{additional confounders} as well.  

\section{Related work}

The Web provides us a giant platform for observing human behavior and allows us to answer various socially relevant questions. With increasing availability of large amount social media data, a significant amount of research efforts have been directed towards understanding human behavior from this data \cite{ruths2014social}. Consequently, research efforts have been able to identify compelling evidence towards the presence of gender inequality in different aspects of social media. 

\noindent\textbf{Gender inequality}: In \cite{wagner2015s}, the authors explore the gender inequalities on Wikipedia and observe that women are portrayed in a starkly different way than men. Gender and racial biases are also observed in online freelance market places~\cite{hannak2017bias}. In \cite{nilizadeh2016twitter}, the authors find that women are underrepresented among the top users on Twitter. A similar effect was documented in the content of online newspapers~\cite{jia2016women}. These online gender biases can have significant economic consequences~\cite{foong2018women}. Sometimes the design of platforms amplify gender differences, suggesting potential points of intervention. On Stack Overflow, a Q\&A platform for programmers, men and women have significant differences in behavior, and platform design choices translate this difference in behavior to a gap in outcomes~\cite{may2019gender}.

\noindent\textbf{Gender perception}: The gaps and disparities described above suggest that users in online communities make assumptions and stereotypes about contributions using socio-demographic features such as age, gender, and ethnicity~\cite{willis2006first}. In \cite{davison2000sex} the authors note that men and women job applicants receive lower ratings for jobs stereotypically held by members of the other gender (e.g. nurses and carpenters). Men are also often rated more competent than women purely based on cues of gender and not content~\cite{fiske2018model}. An analysis of Github, a platform for collaborative programming, using a dataset with self-reported gender identification suggests that contributions of women are accepted more often when their gender is hidden~\cite{terrell2017gender}. 

\noindent\textbf{Gender performance}: Our paper contributes to a growing area of research that describes how gendered performance impacts success and reception. Past work by Otterbacher describes differences in the writing style of IMDB film reviews between men and women~\cite{otterbacher2010inferring}. Otterbacher finds that ``feminine'' reviews are typically ranked as less helpful. More recent work extends the measurement of gendered behavior to visual content creation~\cite{wachs2017men}, music performance~\cite{wang2019gender}, and software engineering~\cite{vedres2019gendered}. A consistent finding across these domains is that when individuals are less successful when they create content with a more feminine style, regardless of their signaled gender~\cite{brooke2019condescending}. 

\section{Conclusions}

In this work, we have quantified the effect of gender signaling on perceived helpfulness of reviews in a large online shopping platform. 

For our analysis, we employed a dictionary based name-to-gender tool to infer gender signals, character-level Convolutional Neural Networks to characterize gender performance, and Mahalanobis matching to measure relationships between these gender features and success. We observed basically no general effect for either gender signaling or performance. Rather, we saw substantial category-specific effects: reviews authored by signaling women are perceived as more helpful in categories like {\myf Toys}, {\myf Movies} and {\myf Beauty} while signaling men receive more kudos for their contributions to categories including {\myf Electronics}, {\myf Kindle}, and {\myf Computers}. 
 In the second dimension of our analysis, we found that in categories like {\myf Electronics} or {\myf Cellphones} gender anonymous reviewers performing as women receive better feedback than signaling women. Similar effects are observed for reviewers performing as men compared with signaling men in categories such as {\myf Books} or {\myf Kindle}. 
 
In the future, it will be interesting to extend our idea to other web platforms and thereby investigate whether gender disclosure and signaling has effects on perceived helpfulness in other domains. The task of inferring gender or gendered behavior from text is a fascinating problem in its own right and demands further inquiry. Finally, we suggest that future work ought to explore the signaling of race and its influence on how content is received online. All together, there is much to do in this line of research.
%%%%%%%%%%%%%%%%%%%%%%%%%%%%%%

\end{document}